# An intelligent Data Delivery Service for and beyond the ATLAS experiment

*Wen Guan*[1,*], *Tadashi Maeno*[2], Brian Paul Bockelman[3], Torre Wenaus[2], Fahui Lin[4], Siarhei Padolski[2], Rui Zhang[1] and Aleksandr Alekseev[5]

[1]University of Wisconsin-Madison, Madison, USA
[2]Brookhaven National Laboratory, Upton, USA
[3]Morgridge Institute for Research, Madison, USA
[4]University of Texas at Arlington, USA
[5]Moscow State U. ; Andres Bello Natl. U. ; Moscow, INR

**Abstract.** The intelligent Data Delivery Service (iDDS) has been developed to cope with the huge increase of computing and storage resource usage in the coming LHC data taking. iDDS has been designed to intelligently orchestrate workflow and data management systems, decoupling data pre-processing, delivery, and main processing in various workflows. It is an experiment-agnostic service around a workflow-oriented structure to work with existing and emerging use cases in ATLAS and other experiments. Here we will present the motivation for iDDS, its design schema and architecture, use cases and current status, and plans for the future.

## 1 Introduction

The ATLAS experiment at the LHC [1] [2] [3] has accumulated about 460 Petabytes of data processed in an internationally distributed Grid infrastructure with around 175 computing centers in more than 40 countries, which is steadily running with the capability of providing about 6M CPU-hours per day. However, when the High Luminosity LHC (HL-LHC) starts its operation circa 2027, the produced data will grow significantly as the luminosity will increase by a factor of 10 beyond the LHC's design value, and ATLAS will be running short of both computing and storage resources. To overcome this challenge, several new workflows have been proposed and developed. For example, the ATLAS Event Streaming Service (ESS) [7][8][9] delivers fine-grained input data to remote computing resources over the network. Another example is the ATLAS Data Carousel, where data processing proceeds as data is staged in from tape storage to minimize the data footprint on disk. Such workflows require close collaboration between the WorkFlow Management system (WFM system) [4][5] and the Distributed Data Management system (DDM system) [6], which can be achieved via the coordination of a high-level service.

---

[*] email: wen.guan@cern.ch

The iDDS system has been developed to orchestrate WFM and DDM systems in order to optimize resource usage in various workflows. It dynamically transforms and delivers data to let computing resources process data promptly, decoupling data pre-processing, delivery, and main processing in each workflow and allowing them to run asynchronously. The main functions of iDDS are:

- On-demand data transformation: To transform source data on the storage side to the format optimal for delivery to the consumer and subsequent processing, to minimize the network traffic and reduce usage of local disks or caches.
- Data delivery with optimal granularity: To partition (as appropriate) data to an optimal granularity for delivery, while preserving effective data caching.
- Intelligent orchestration: To orchestrate WFM and DDM systems to execute tasks with optimized resource usage by managing dataflow and workflow based on knowledge of the data-driven execution graph, data locality and status, available caches, available processing workers, and other dynamic workflow characteristics.

## 2 iDDS architecture

iDDS consists of a general RESTful head service to receive requests from clients, and several daemons to process the requests. The schematic view of the iDDS architecture is described in Figure 1. The RESTful head service authenticates users, registers and queries requests, and provides an interface to look up data collections or their contents associated with the requests. There are five types of daemons: Clerk, Marshaller, Transformer, Carrier, and Conductor. The Clerk manages requests and converts them to Workflow objects. The Marshaller manages directed acyclic graphs (DAGs) and splits Workflow objects to Work objects. One Work object corresponds to one data transformation, and one Workflow object represents a group of Work objects and their relationships. The Transformer takes care of association between input and output data, interacts with the DDM system if necessary, and creates Processing objects to transform data. The Carrier submits Processing objects to the WFM system and periodically checks their status. The Conductor checks availability of output data and sends notifications to data consumers to trigger subsequent processing.

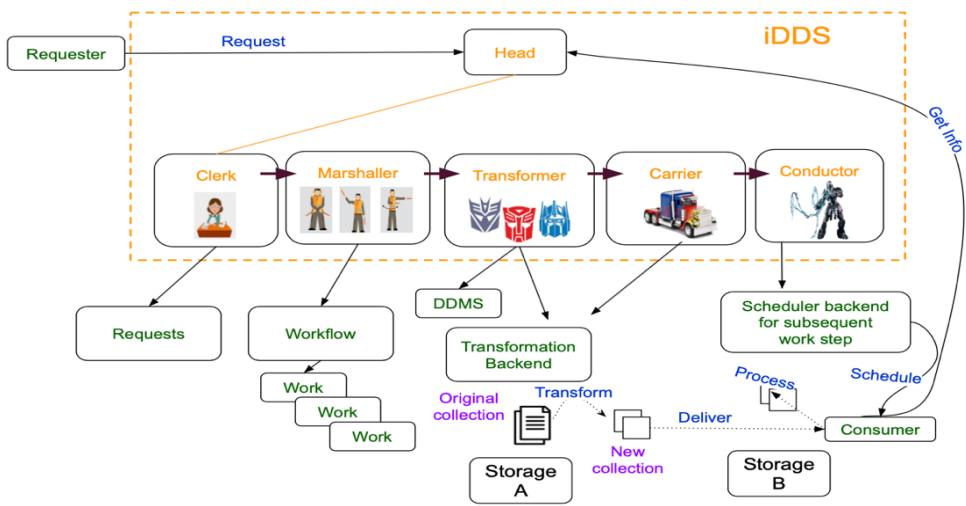

**Fig. 1.** A schematic view of the iDDS architecture

Clients define Workflows, which are serialized to json-based requests, and submit the requests to the RESTful head service. The requests are deserialized on the server side to be passed to iDDS daemons as shown in Figure 2.

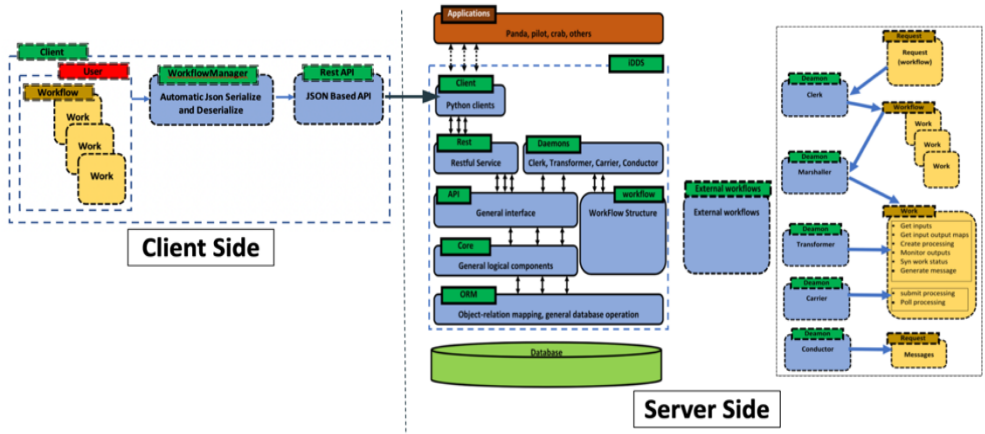

**Fig. 2.** Communication between client and iDDS.

The DG (Directed Graph) workflow management in iDDS not only supports DAG (Directed Acyclic Graph), but also supports graphs with cycles. A DG is represented as a Workflow object which is composed of multiple Work template objects and their relationship with condition branches, as shown in Figure 3. A Work template is a placeholder to generate new Work objects by assigning values for pre-defined parameters. When a Work is terminated, all associated Condition branches will be evaluated and new Work objects can be generated from their following Work templates, with newly assigned values for pre-defined parameters.

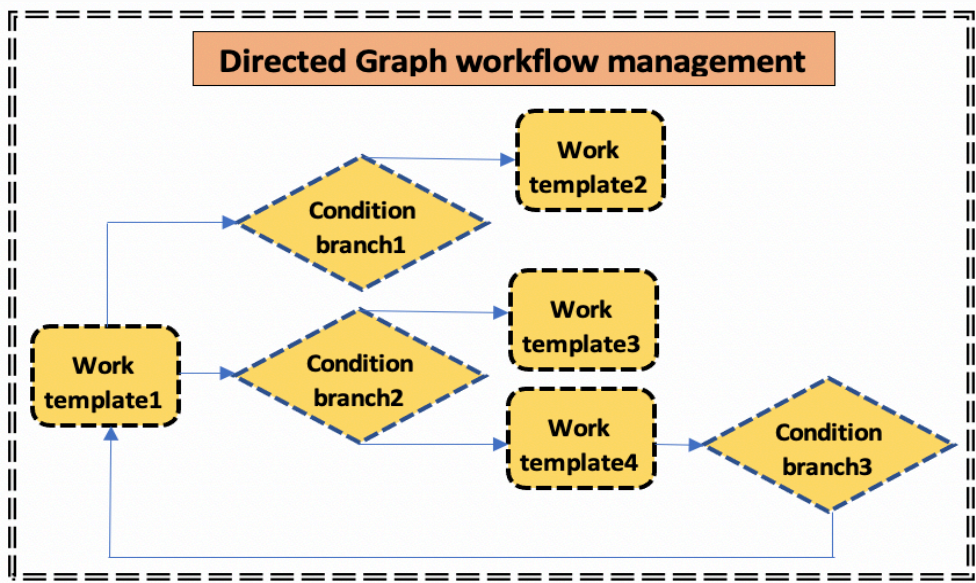

**Fig. 3.** Directed Graph workflow management in iDDS.

# 3 Use Cases

## 3.1 Optimization of ATLAS Data Carousel

The idea of the ATLAS data carousel is to increase the usage of less expensive tape storage relative to expensive disk [10][11]. The first data carousel implementation worked with coarse dataset-level data granularity due to constraints in the WFM and DDM systems, which caused significant overhead before processing the data and required big disk pools to cache the data during the whole processing period. An optimally implemented data carousel starts processing data as soon as it appears from tape, not when most of the input data is ready. iDDS brought the implementation much closer to this optimum than the first implementation with coarser granularity. In the current implementation, iDDS has added the capability to the WFM system to work with fine-grained file-level data. Input data is incrementally processed based on more detailed knowledge on the status of input data, to reduce the overhead and get rid of redundant data transfers and caching. Processed data is released from the cache promptly with similarly fine granularity, such that the full workflow minimizes the input data footprint on disk. iDDS has been integrated with the ATLAS computing system since mid 2020 and has been used for bulk data reprocessing campaigns. The status of data reprocessing with iDDS is shown in Figure 4 and Figure 5.

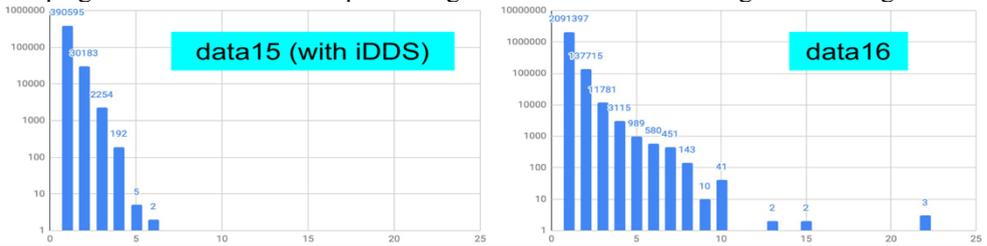

**Fig. 4.** Job attempt times comparation with and without iDDS. iDDS reduces a lot of job attempts.
(X: Number of attempts, Y: Number of jobs.)

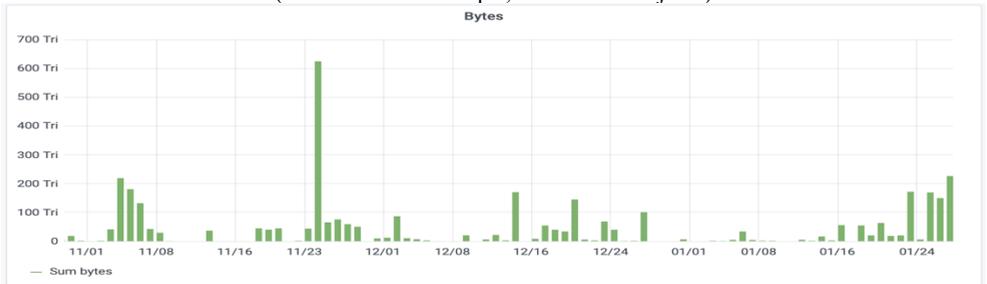

**Fig. 5.** Status of data reprocessing with data carousel + iDDS
(The number of bytes processed per day for last 90 days)

## 3.2 Hyperparameter Optimization Service

Machine learning is becoming an important tool for data analysis in ATLAS and the wider community. A hyperparameter is a parameter to control the training process in machine learning. Hyperparameter Optimization (HPO) is to choose a set of optimal hyperparameters for a machine learning algorithm, and can be resource intensive. iDDS provides a fully automated platform for HPO on top of geographically distributed GPU resources among the grid, HPC, and clouds, such that large scale resources can be applied to large HPO tasks. Figure 6 shows how iDDS implements the HPO workflow, where iDDS centrally scans the search space using advanced optimization algorithms to generate hyperparameter points, while hyperparameter points are asynchronously evaluated on remote GPU resources. The training results with those hyperparameter points are reported back to iDDS for further optimization of the search space, and to generate a new round of hyperparameter points. Eventually users get the best hyperparameter point and resultant trained models after all iterations are done. This service is up and running for ATLAS machine learning users. The HPO service is experiment-agnostic by design, so that it should be easy to use it outside of ATLAS, but that has not been tried as yet.

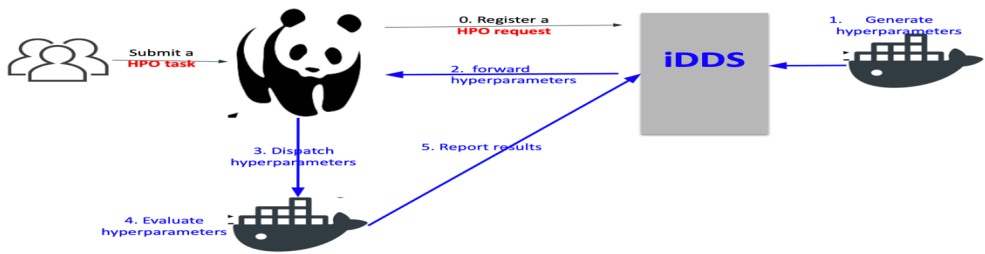

**Fig. 6.** The structure of iDDS Hyperparameter Optimization service.

## 3.3 Directed Graph based workflows

### 3.3.1 Rubin Observatory exercise

The Rubin Observatory (LSST) [12] exercise is an ongoing activity to evaluate PanDA as both a workflow and workload management system. A workflow graph is dynamically generated by Rubin middleware for each payload submission and includes, among others, a set of dependencies for each individual job that must be satisfied before the job could be processed. A single workflow can consist of a hundred thousand jobs forming the vertexes of a DAG. It is the first use case of the DG-based workflow support in iDDS. Every workflow is mapped to sequentially concatenated Work objects in iDDS. iDDS also allows Work objects to be incrementally released based on messaging, in order to avoid long waiting in each Work.

### 3.3.2 Active Learning

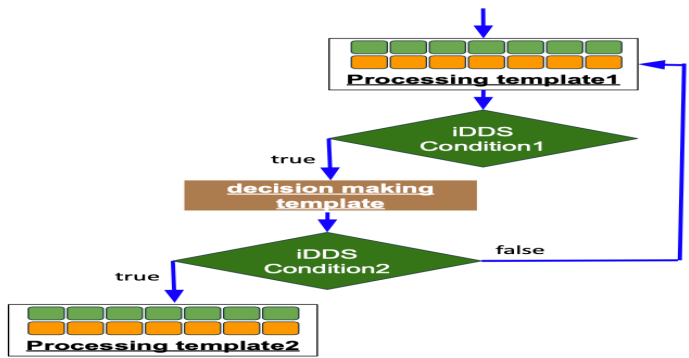

**Fig. 7.** The structure of iDDS Active Learning service.

Active Learning is another use case of DG based workflow support, developed initially for ATLAS. There are two types of Work objects: one for processing and the other for decision making. The decision making Work object takes output data from the upstream processing Work object to provide hints to the downstream processing Work object.

In Active Learning, as shown in Figure 7, Work templates are defined as placeholders of the processing and the decision-making Work objects, with pre-defined parameters. When a Work completes, its associated Condition branching objects will be evaluated, to check whether to trigger next processing, which processing to be triggered, and what new values for next processing's pre-defined parameters.

This workflow is currently a prototype. It is being evaluated inside iDDS and is being integrated with PanDA for ATLAS usage.

## 4 Summary and Outlook

iDDS has been developed to support various emerging use cases in ATLAS and other experiments. It has already been in production for data carousel and hyperparameter optimization services in ATLAS, and is being evaluated for Rubin Observatory (LSST) [12]. The workflow-oriented structure of iDDS makes it straightforward to add support for new use cases. Current priorities are to improve the user experience of the client API and CLI tool, documentation, and monitoring. We anticipate adding more use cases in multiple experiments.

This work was supported by the National Science Foundation under Cooperative Agreement OAC-1836650.